\documentclass[aps,prl,superscriptaddress,onecolumn,normalem]{revtex4-1}
\usepackage{amsmath,amssymb,bm,graphicx,epsfig,psfrag,ulem,color,natbib, comment}
\usepackage{hyperref}

\newcommand\Rey{\mbox{\textit{Re}}}  % Reynolds number
\newsavebox{\astrutbox}
\sbox{\astrutbox}{\rule[-5pt]{0pt}{20pt}}

\newcommand\etal{\mbox{\textit{et al.}}}

\newcommand{\bs}{\mathbf {s}}
\newcommand{\br}{\mathbf {r}}
\newcommand{\bx}{\mathbf {x}}

\newcommand{\bv}{\mathbf {v}}

\begin{document}

\title{Thermal counterflow in a periodic channel with solid boundaries}

\author{Andrew~W.~Baggaley}
\email{andrew.baggaley@gla.ac.uk}
\affiliation{School of Mathematics and Statistics, University of
Glasgow, Glasgow, G12 8QW, United Kingdom}
\author{Jason Laurie}
\email{jason.laurie@weizmann.ac.il}
\affiliation{Department of Physics of Complex Systems, Weizmann Institute of Science, Rehovot, 76100, Israel}

%\date{\today}

\begin{abstract}
We perform numerical simulations of finite temperature quantum turbulence produced through thermal counterflow in superfluid
$^4$He, using the vortex filament model. We investigate the effects of solid boundaries along one of the Cartesian directions, assuming a laminar normal fluid with a Poiseuille velocity profile, whilst varying the temperature and the normal fluid velocity. We analyze the distribution of the quantized vortices, reconnection rates and quantized-vorticity production as a function of the wall-normal direction.  We find that the quantized vortex lines tend to concentrate close to the solid boundaries with their position depending only on temperature and not on the counterflow velocity.  We offer an explanation of this phenomenon by considering the balance of two competing effects, namely the rate of turbulent diffusion of an isotropic tangle near the boundaries and the rate of quantized-vorticity production at the centre. Moreover, this yields the observed scaling of the position of the peak vortex line density with the mutual friction parameter.  Finally we provide evidence that upon the transition from laminar to turbulent normal fluid flow, there is a dramatic increase in the homogeneity of the tangle, which could be used as an indirect measure of the transition to turbulence in the normal fluid component for experiments.
\end{abstract}

%Something or other

\maketitle

\section{Introduction}
%%What is QT?

Thermal counterflow is a form of turbulence unique to the two-fluid description of finite temperature superfluid helium. Here, a viscous normal fluid coexists with an inviscid superfluid; the relative densities of the two fractions are temperature dependent. In the viscous normal fluid, vorticity is unconstrained with rotational motion occurring over a wide range of scales and intensities. In contrast, the superfluid component is characterized by irrotational fluid motion around stable topological defects--atomically thin quantized vortex lines with fixed core size of approximately $a \simeq 10^{-8}~\mathrm{cm}$ for superfluid helium-4, which is many orders of magnitude smaller than the typical experimental system size.  Crucially, quantum restrictions lead to a fixed  circulation around each quantized vortex line (integer multiples of $\Gamma = h/m$, where $h$ is Planck's constant and $m$ is the mass of a helium atom).  
These systems are as close as one can get to producing the idealized vortices of classic textbooks \cite{saffman1992vortex} within the laboratory.

%%What is counterflow?
Counterflow is a mechanism for heat transport within quantum turbulence, the relative motion of the two fluid components (normal and superfluid) prevents the formation of `hot spots', which subsequently inhibit the formation of bubbles during rapid vaporization of liquid helium below the critical temperature for superfluidity. However, as noticed by Vinen~\cite{Vinen1,Vinen2,Vinen3,Vinen4}, in an important series of papers, there is a limit to the 
thermal counterflow's ability to perfectly equilibrate temperature in the system. If one takes a channel filled with superfluid helium, and heats one end, then counterflow is created with the normal fluid flowing away from the heater, and the superfluid (carrying zero entropy) moving towards the heater.
If the applied heat flux exceeds a critical value, the ideal flow of the superfluid
is lost, and  a disordered arrangement of quantized vortex filaments, often described as a vortex tangle, is created. The motions of the superfluid component induced by the tangle is often referred to as quantum turbulence \cite{VinenNiemela,Skrbek12}, and provides a rich tapestry for those interested in vortex dynamics and the mechanisms of turbulence. 

Counterflow is a unique form of turbulence, in the sense that there is no classical analog.  However, there is still much merit in its study, even if one is interested in classical phenomena. 
For example, Skrbek~\etal~\cite{SkrbekPRE2003} pointed to the fact that the efficiency of turbulent heat transport in counterflow is similar to that of turbulent thermal convection in a classical fluid. 
Recent advances in the visualization of the normal fluid velocity~\cite{Guo2010} promise to revolutionize our knowledge of the flow of the normal fluid component, and in particular, the transition from the laminar to the turbulent state.
One would expect that these advances would offer valuable insight into the transition to turbulence in classical fluids. Moreover, the decay of counterflow turbulence has been shown to have a classical nature~\cite{SkrbekPRE2003}, why this is the case is still an outstanding question in the field.

Driven by modern advances in computational power, a new generation of numerical studies have built on the pioneering work of Schwarz~\cite{Schwarz1988} to offer new insights into the problem of counterflow. Barenghi and Skrbek~\cite{Barenghi2007} used numerical simulations to offer a solution to the puzzle of the initial stages of decaying counterflow, where experimental data suggested that the vortex line density could increase after the heater had been switched off. Adachi~\etal~\cite{Adachi2010} showed, using the full Biot-Savart law, that some statistical properties of the tangle compared well to results from experimental studies. In a similar study~\cite{Adachi2011}, the velocity statistics of the superfluid turbulent flow  were also shown to agree with experimental observations. More recently, Baggaley~\etal~\cite{Sherwin2012} probed the structure of the quantum turbulence generated through counterflow, and showed distinct differences when compared to quantum turbulence generated in a classical manner, i.e. through mechanical `stirring' of the system.

What is common with all of these recent studies, is the use of periodic boundary conditions or an analysis of the system far away from the influence of solid boundaries. An older study by Aarts and de Waele~\cite{Aarts1994} considered the effect of boundaries in both a channel and a pipe, however the computational constraints of the time limited the depth of the study, in comparison to what is now achievable. Galantucci~\etal~\cite{Galantucci2011}, studied two-dimensional quantum turbulence in a periodic channel with solid boundaries under the influence of a Poiseuille normal fluid flow. Their results detailed the steady state vortex density profiles across the channel with regards to several {\it ad hoc} vorticity injection mechanisms. 

In this article, we consider the three-dimensional generalization of the latter.  However, in our setup vorticity does not need to be artificially injected into the system as a three-dimensional quantum tangle is naturally generated through the normal fluid. We implement the vortex line density visualization technique outlined in a paper from one of the present authors~\cite{BaggaleyLaizet} in order to investigate the effect of the solid boundaries upon the superfluid flow for various temperatures and counterflow velocities. The focus of the previous work was to investigate if turbulence in the normal fluid could support a higher vortex line density for a given counterflow velocity, and ultimately how such an increase in the vortex line density could arise. Only two temperatures were studied, each for a single counterflow velocity; no careful analysis of the profile of the vortex line density across the channel was performed. 

In this manuscript, we perform a much more detailed investigation of the laminar regime over a wider range of temperatures and counterflow velocity. We identify universal flow behavior with respect to temperature and the counterflow velocity for the vortex line density, reconnection rate and quantized vorticity production profiles across the channel. In addition, we extend the work of \cite{BaggaleyLaizet}, which used a frozen snap-shot of Navier-Stokes turbulence as a model for a turbulent normal fluid profile. By applying a similar technique, we discover universal characteristics of the vortex line density profile that were also observed in the laminar regime, although with a somewhat enhanced homogeneity of the quantum turbulence tangle. More significantly, we point out the fact that this increase in the homogeneity of the tangle could be measured experimentally, allowing for the transition to turbulence of the normal fluid component to be probed indirectly.  Where possible, we draw direct comparisons of our results to previous numerical and experimental studies.

\section{The Numerical Setup}

We numerically simulate counterflow turbulence using the well-known vortex filament model of Schwarz~\cite{Schwarz1985}.  This entails modeling the motion of the quantized vortex lines as one-dimensional space curves, $\bs(\xi,t)$, which evolve according to
\begin{equation}
\frac{{\rm d}{\bf s}}{{\rm d}t}=\bv_s+\alpha \bs' \times (\bv_n-\bv_s)
-\alpha' \bs' \times \left[\bs' \times \left(\bv_n-\bv_s\right)\right],
\label{eq:Schwarz}
\end{equation}

\noindent
where $t$ is time, $\alpha$ and $\alpha'$ are known
temperature dependent friction
coefficients \cite{Donnelly1998}, $\bs'={\rm d}\bs/{\rm d}\xi$ is the unit
tangent vector at the point $\bs$, $\xi$ is arc length, and
$\bv_n$ is the normal fluid velocity at the point $\bf s$.

The velocity of the superfluid component can be further decomposed, such that 
$\bv_s=\bv_s^{\rm si}+\bv_s^{\rm ext}$. 
Here $\bv_s^{\rm si}$ represents the self-induced velocity
of the vortex line at the point $\bs$, given by
the Biot-Savart law~\cite{saffman1992vortex}
\begin{equation}
\bv_s^{\rm si} (\bs,\,t)=
\frac{\Gamma}{4 \pi} \oint_{\cal L} \frac{(\br-\bs) }
{\vert \br - \bs \vert^3}
\times {\rm {\bf d}}\br,
\label{eq:BS}
\end{equation}
\noindent
where $\Gamma=9.97 \times 10^{-4}~\rm cm^2/s$
(in $^4$He) and the line integral
extends over the entire vortex configuration ${\cal L}$. 
$\bv_s^{\rm ext}$ is an `external' superflow due to the normal fluid and the conservation of mass, given by $\bar{v}_s^{\rm ext}\equiv\left\langle \bv_s^{\rm ext}\cdot {\bf e}_x\right\rangle_V=\rho_n \bar{v}_n/\rho_s$,
where $\bar{v}_n\equiv\left\langle \bv_n\cdot {\bf e}_x\right\rangle_V$ is the mean flow rate of the normal fluid along the channel. $\left\langle\cdot\right\rangle_V$  denotes the volume average, and  $\rho_n$ and $\rho_s$ are the relative densities of the normal and superfluid components respectively.

The normal fluid velocity $\bv_n$, satisfies the Navier-Stokes equations coupled to the vortex filament model via a mutual friction term ${\bf F}_{ns}$:

\begin{eqnarray}\label{eq:NavierStokes}
\frac{\partial \bv_n}{\partial t} + \left(\bv_n\cdot\nabla\right) \bv_n &=& -\frac{1}{\rho_n}\nabla P + \nu \Delta \bv_n + \frac{1}{\rho_n}{\bf F}_{ns}+ {\bf f},\\
\nabla\cdot \bv_n &=& 0,
\end{eqnarray}
where $P$ is the pressure, $\nu$ is the kinematic viscosity of helium-4, and {\bf f} is an external forcing.

The explicit expression of the mutual friction term, $\mathbf{F}_{ns}$ arising from drag of the quantized vortex lines is given by~\cite{Kivotides2000}
\begin{equation}\label{eq:mutualfriction}
{\bf F}_{ns}(\bx) =  \oint_{\mathcal{L}} D\left(\left[{\bs}' \times \left( \bv_n - \dot{\bs}\right)\right] + \left(D'-\rho_n\Gamma\right)\left\{{\bs}' \times \left[{\bs'}\times \left( \bv_n - \dot{\bs}\right)\right]\right\}\right)\delta(\bx - \bs)\ {\rm d}\xi({\bf s}),
\end{equation}
where $D$ and $D'$ are the standard temperature dependent mutual friction coefficients given in Table~\ref{tab:parameters}.  The mutual friction term (\ref{eq:mutualfriction}) constitutes a line integral over the entire vortex tangle $\mathcal{L}$ which implies that the normal fluid only feels the mutual friction precisely along the quantized vortex lines.
 
Ideally, both coupled Eqs.~(\ref{eq:Schwarz}) and (\ref{eq:NavierStokes}) should be solved simultaneously, however the computational expense of such a numerical scheme prevent us at this moment in time. Instead, we consider counterflow velocities that are sufficiently small to ensure that the normal fluid flow remains laminar (the so-called TI state~\cite{Tough1982}). Therefore, we forgo the expense of solving the Navier-Stokes equations and instead consider a stationary normal fluid velocity field of the form of a laminar planar Poiseuille profile
\begin{equation}
\label{eq:profile}
\bv_n(y)= \left(1-\frac{y^2}{h^2}\right) U_c \, {\mathbf{e}}_x,
\end{equation}
where $h=\mathcal{D}_{y}/2$ is half of the channel width and $\mathbf{e}_x$ is the unit vector along the $x$-direction.   In making this assumption, not only do we assume that the normal fluid Reynolds number, $\Rey$, is sufficiently small, but also that the presence of the quantum tangle does not destabilize the normal fluid flow. Whilst a detailed experimental investigation of the normal fluid profile in the laminar regime has not yet been performed, it is worth noting that an approximately parabolic velocity profile in the normal fluid has been observed for small applied heat fluxes~\cite{Vinen2014rev}.

\begin{figure*}
\begin{center}
\includegraphics[width=0.36\textwidth]{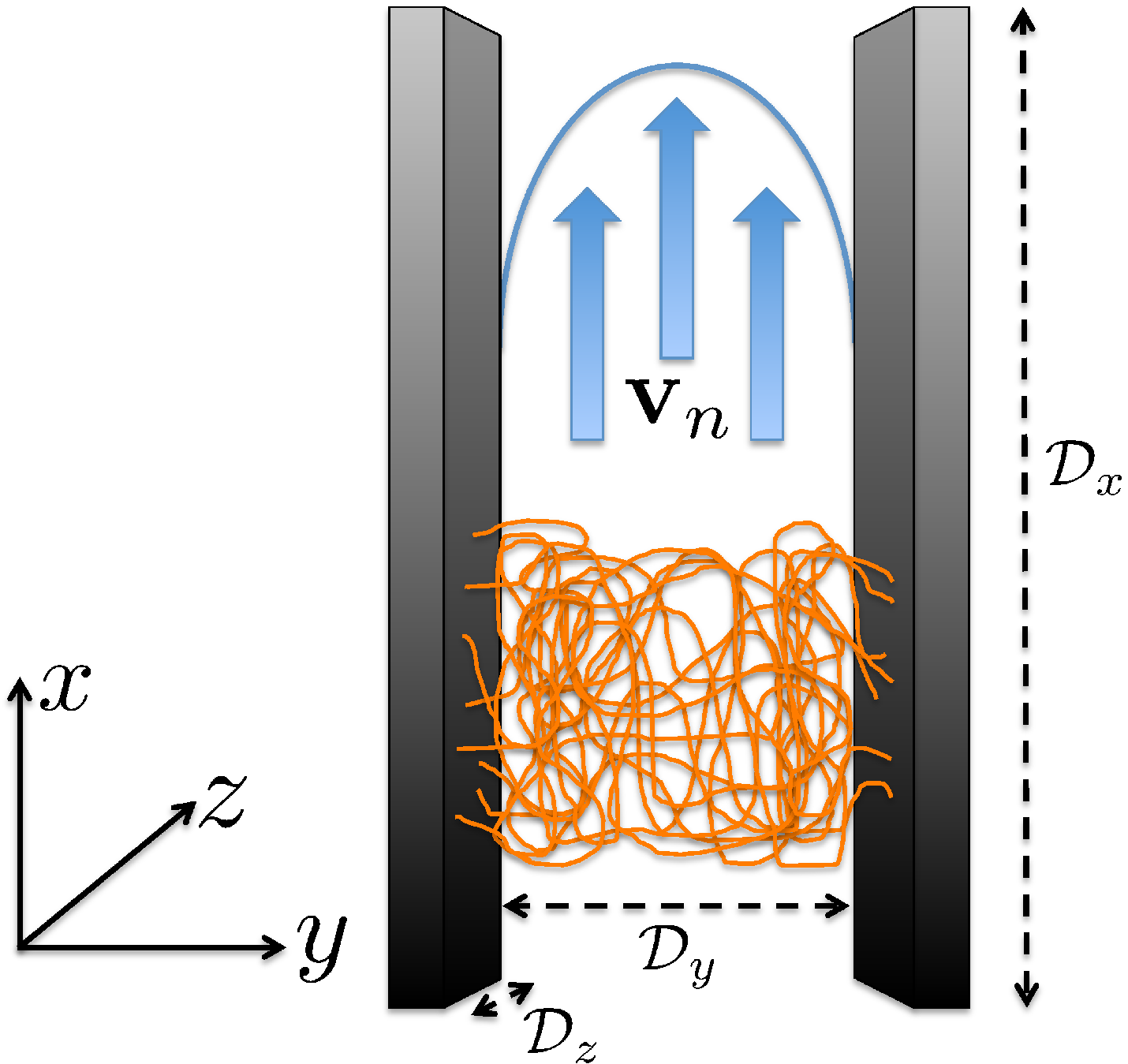}
\hfill
\includegraphics[width=0.17\textwidth]{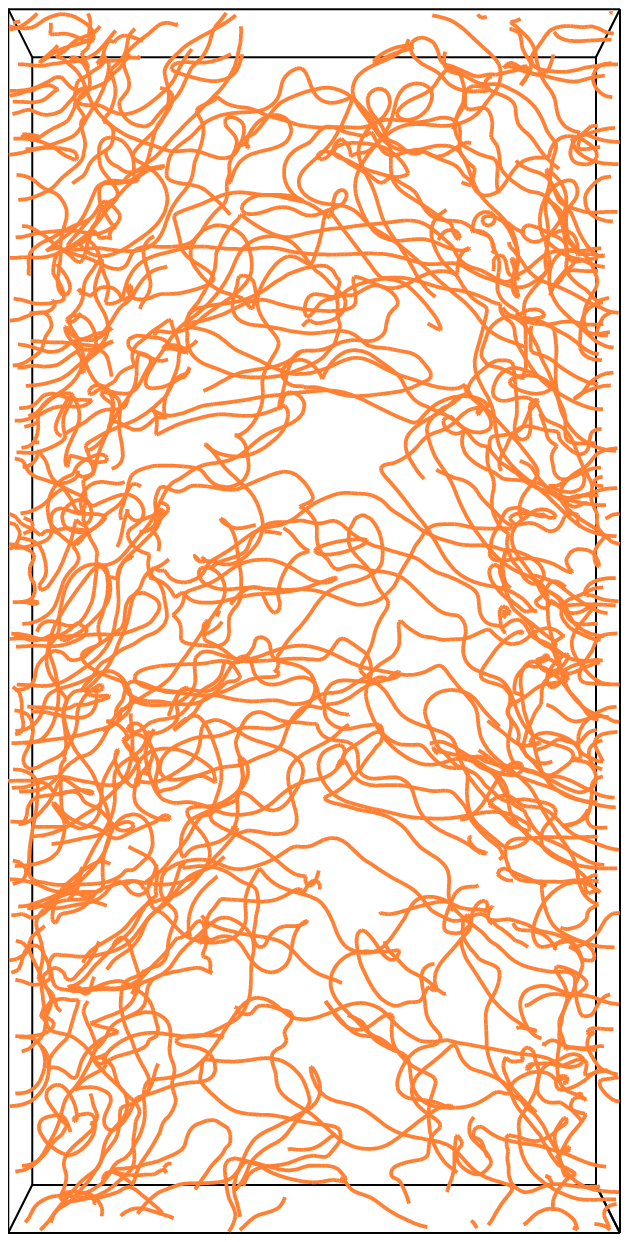}
\hfill
\includegraphics[width=0.17\textwidth]{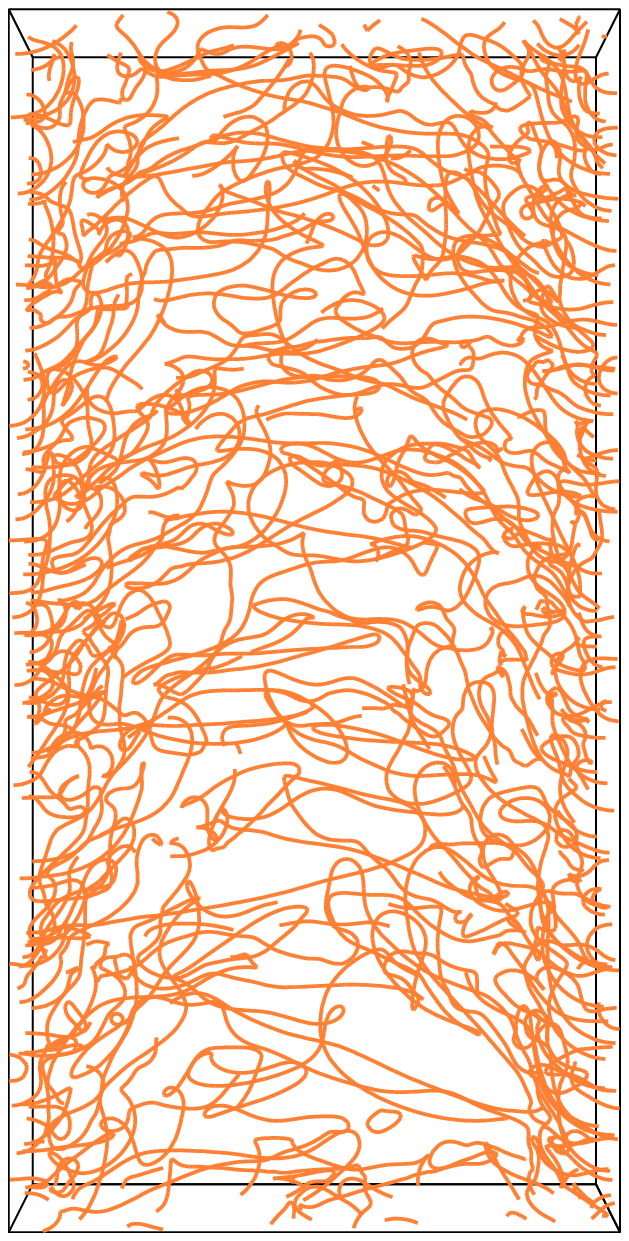}
\hfill
\includegraphics[width=0.17\textwidth]{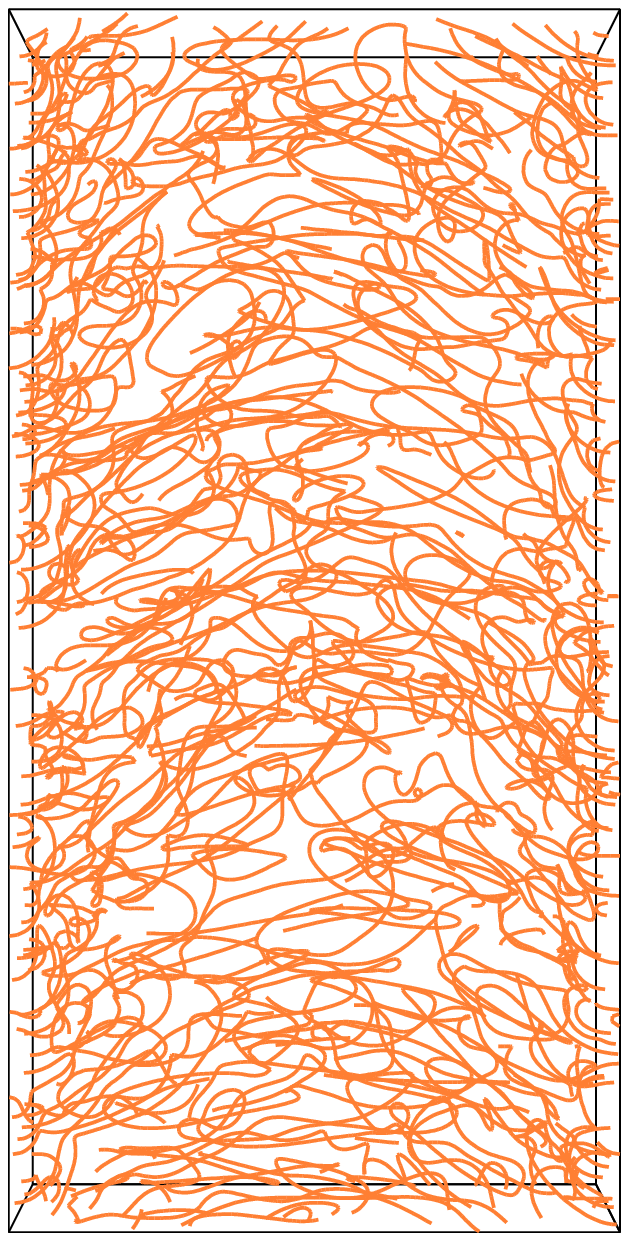}
\caption{(Far left) Schematic of the Numerical Box. We have solid boundaries in the $y$-direction, and periodic in $x$ and $z$.  There is a normal laminar fluid flow with a  Poiseuille profile (Eq.~(\ref{eq:profile})) directed along the $x$-direction which forces a quantum vortex tangle depicted in orange. Snapshots of the vortex tangle during the steady-state regime with $T=1.3 ~\rm K$, $\bar{v}_{ns}=1.5~\rm cm/s$ (second left), $T=1.6 ~\rm K$, $\bar{v}_{ns}=1.13~\rm cm/s$ (second right) and $T=1.9 ~\rm K$, $\bar{v}_{ns}=0.90~\rm cm/s$ (far right).}
\label{fig1}
\end{center}
\end{figure*}

All calculations are performed in a cuboid of size $\mathcal{D}_x \times \mathcal{D}_y \times \mathcal{D}_z=0.2~\rm cm \times 0.1~\rm cm \times 0.1~\rm cm$, with solid boundaries in the $y$-direction and periodic boundaries in $x$ and $z$ (see Fig.~\ref{fig1}).

%%%%%%%%%%%%%%%%%%%%%%%%%%%%%%%%%%%%%%%%%%%%%%%%%%
%\begin{figure}
%\begin{center}
%\includegraphics[width=0.35\textwidth]{Box.eps}
%\caption{Schematic of the Numerical Box. We have solid boundaries in $y$, and periodic in $x$ and $z$.  There is a normal laminar fluid flow with a  Poiseuille profile (Eq.~\eqref{eq:profile}) directed along the $x$-direction which forces a quantum vortex tangle depicted in orange.}
%\label{fig1}
%\end{center}
%\end{figure}
%%%%%%%%%%%%%%%%%%%%%%%%%%%%%%%%%%%%%%%%%%%%%%%%%%

The numerical techniques to discretize the vortex lines
into a number of points $\bs_j$ ($j=1, \cdots N$) held
at minimum separation $\Delta\xi/2=5 \times 10^{-4}~$cm, de-singularize the Biot-Savart integrals, 
evaluate $\bv_s$ using a tree-method (with critical opening angle $0.3$), and 
algorithmically model vortex reconnections when vortex lines
come sufficiently close to each other, are all
described in previous 
papers~\cite{BaggaleyFluc,BaggaleyRecon}.
Integration in time is achieved using a third order Runge-Kutta scheme with a
timestep $\Delta t=5 \times 10^{-5}~$s.

%%%%%%%%%%%%%%%%%%%%%%%%%%%%%%%%%%%%%%%%%%%%%%%%%%
%%%%%%%%%%%%%%%%%%%%%%%%%%%%%%%%%%%%%%%%%%%%%%%%%%

We work at a number of experimentally feasible temperatures given in Table~\ref{tab:parameters}.
%;  $T=1.9~\rm K$ (corresponding to $\alpha=0.206$ and $\alpha'=0.0083$), $T=1.6~\rm K$ ($\alpha=0.098$ and $\alpha'=0.016$), and $T=1.3~\rm K$ ($ \alpha=0.036$ and $\alpha'=0.014$).
The counterflow direction is parallel to the elongated $x$-axis. The choice of solid boundaries in only one direction is favourable for two reasons, firstly it allows us to average over two periodic directions to obtain much better statistics of the tangle. Secondly, the majority of studies of turbulent channel flow~\cite{moseretal99} use Direct Numerical Simulations (DNS) of the Navier-Stokes equations for a flow between two boundaries in a periodic channel, therefore consistency with the boundary conditions will allow a direct comparison.

Our solid boundaries are idealized with regards to being completely smooth.  In reality this is not the case, with experimental apparatuses being constructed of materials with rough walls in comparison.  This would have the effect of introducing inhomogeneities in the flow and emphasizing the boundary layer effect of the normal fluid. However, Schwarz~\cite{Schwarz1985} showed only a relatively modest velocity is required to de-pin vortices and we find good agreement between our results and those of experimental studies. The effects of the solid boundaries on the inviscid superfluid is modeled by the presence of two {\it image} vortex tangles arising from the reflection of the original tangle through the two boundaries, plus four periodic copies. This results in the desired free-slip boundary conditions for the superfluid flow. Consequently, we assume that the profile of the external superflow can be suitably taken as a constant profile across the channel. Conversely, the normal fluid flow must satisfy the no-slip boundary condition, therefore an appropriate model for the laminar normal fluid, in which the velocity vanishes at the boundary, is the Poiseuille profile~(\ref{eq:profile}).

\begin{table}
  %\begin{center}
%\def~{\hphantom{0}}
  \caption{Values of the parameters used in our simulations; taken from \cite{Barenghi1983}. Here $T$ denotes the temperature of the system; $\nu$ the kinematic viscosity; $\rho_n$ the density of the normal fluid component and $\rho_s$ the density of the superfluid component; $\alpha$ and $\alpha'$ are the non-dimensionalized mutual friction coefficients of the vortex filament model~(\ref{eq:Schwarz}); and $D$ and $D'$ are the standard mutual friction coefficients in which $\alpha$ and $\alpha'$ are defined.  \label{tab:parameters}}
  %\begin{tabular}{cccccccc}
  \begin{tabular}{lllllllll}
  \hline\noalign{\smallskip}
      $T{~[\rm K]}$ & $\nu~[\rm cm^2/ s]$ &$\rho_n~[\rm g /cm^{3}]$& $\rho_s~[\rm g/cm^{3}]$ & $\alpha$   &   $\alpha'$ & $D~[\rm g/cm\, s]$ & $D'~[\rm g /cm\, s]$ \\ 
      \noalign{\smallskip}\hline\noalign{\smallskip}
       $1.3$ &  $1.05\times 10^{-4}$ & $6.52\times 10^{-3}$ & $1.39\times 10^{-1}$& $3.6\times 10^{-2}$ & $1.4\times 10^{-2}$ & $5.93\times 10^{-6}$ & $4.16\times 10^{-6}$\\
       $1.6$ &  $9.00\times 10^{-5}$ & $2.36\times 10^{-2}$ & $1.22\times 10^{-1}$  & $9.8\times 10^{-2}$ & $1.6\times 10^{-2}$ & $2.79\times 10^{-5}$ & $1.96\times 10^{-5}$\\
       $1.9$ & $9.26\times 10^{-5}$  & $6.10\times 10^{-2}$ & $8.44\times 10^{-2}$ & $2.1\times 10^{-1}$ & $8.3\times 10^{-3}$ & $4.71\times 10^{-5}$ & $1.35\times 10^{-4}$\\
       \noalign{\smallskip}\hline
         \end{tabular}

%  \end{center}
\end{table}

The relative motion of the normal and superfluid velocity fields is commonly denoted by $\mathbf{v}_{ns}\equiv \mathbf{v}_n - \mathbf{v}_s$ and is known as the counterflow velocity. The mean value of $\bv_n$, from Eqs.~(\ref{eq:profile}) is then given by ${\bar v}_n=2 U_c/3$; and so $\bar{v}_{ns}=\bar{v}_n+\rho_n \bar{v}_n/\rho_s$.   Our simulations are performed within the range $0.478 \,{\rm cm/s} \le \bar{v}_{ns} \le 1.5 \, {\rm cm/s}$, with values chosen to ensure comparable vortex line densities across the temperature range.

\section{Results}

We perform twelve simulations using the Poiseuille profile, at four different values of $\bar{v}_{ns}$ for each of the three temperatures used in this study.  The corresponding Reynolds numbers, $\Rey$, for the Poiseuille flows are all under ${\Rey}\leq 1069$, far below the critical Reynolds number for classical turbulent channel flow ${\Rey}_c\simeq 5772$~\cite{Orszag1971}. In all the simulations we initialize the system with 50 randomly oriented loops of radius $7.0 \times 10^{-3}~{\rm cm}$, confined to the computational domain.   We let each simulation evolve in time such that a statistical steady state is achieved, checked by observing stationarity of the vortex line density.   Snapshots of the superfluid vortex tangles in steady state conditions for each of the temperatures for a specific counterflow velocity can be seen in Fig.~\ref{fig1}.  Visually one can see a marked concentration of vortex lines close to the solid boundaries, however it is only through a detailed statistical analysis of the vortex configuration in the steady state regime that the structure of the vortices is truly exposed.

\subsection{Steady state vortex line density}

One of the most fundamental quantities that characterizes quantum turbulence is the vortex line density $L\equiv \Lambda/V$, where $\Lambda(t) = \int_{\mathcal{L}}\,  {\rm d}\xi$ is the total vortex line length constituting of the integration of the arc length $\xi$ taken over the whole vortex line configuration $\mathcal{L}$ and then divided by the domain volume $V$.

\begin{figure}
\begin{center}
\includegraphics[width=0.42\textwidth]{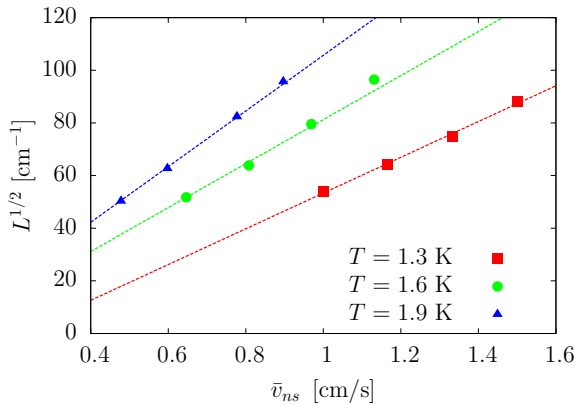}
\caption{Steady state vortex line densities plotted as a function of the counterflow velocity $\bar{v}_{ns}$. Squares (red) denote the $T=1.3~\rm K$ simulations, circles (green) $T=1.6~\rm K$, and triangles (blue) $T=1.9~\rm K  $. Lines of best fit corresponding to the relationship $L^{1/2}=\gamma(T) \left(\bar{v}_{ns}-v_0\right)$ are plotted; fitting parameters $\gamma=67.9~\rm s/cm^2$, $v_0=0.21~\rm cm/s$ ($T=1.3~\rm K$), $\gamma=83.6~\rm s/cm^2$, $v_0=2.8\times 10^{-2}~\rm cm/s$ ($T=1.6~\rm K$) and $\gamma=105.7~\rm s/cm^2$, $v_0=1.5\times 10^{-4}~\rm cm/s$ ($T=1.9~\rm K$).}
\label{fig2}
\end{center}
\end{figure}

Vinen~\cite{Vinen3,Vinen4} introduced a phenomenological
model for a homogeneous vortex tangle, where the evolution of the vortex line density $L(t)$ was described through a balance of vortex production due to the imposed counterflow and decay due to dissipation.  In steady state conditions, where ${\rm d}L/{\rm d}t=0$, Vinen's equation yields the relation
\begin{equation}\label{eq:gammarelation}
L^{1/2} = \gamma(T) (\bar{v}_{ns}-v_0),
\end{equation}  
where $v_0$ is an additional fitting parameter, and $\gamma$ is a temperature dependent parameter.  The measurement of $\gamma$ and the verification of relation~(\ref{eq:gammarelation}) from experiments and numerical simulations has been quite robust and universal with only minor discrepancies arising from different setups.  Likewise, we check this relation for our numerical configuration.  We observe in Fig.~\ref{fig2} excellent linear scalings with regards to the counterflow velocity $\bar{v}_{ns}$ for all three temperature regimes.  The values we obtain for $\gamma$ are comparable to those obtained in the numerical work of Adachi~\etal~\cite{Adachi2010}  (with triply periodic boundary conditions and a constant counterflow velocity), as well as the experimental study of Tough~\cite{Tough1982} where it is believed that the normal fluid is laminar. In contrast, recent experimental studies (see~\cite{Babuin12}) consider superflow as opposed to counterflow, report higher values of $\gamma$, however here the normal fluid is probably turbulent. 

For the critical velocities $v_0$, we find values much smaller than observed experimentally~\cite{Chagovets2008}, however they are consistent with other numerical studies. The discrepancy with experiments is probably due to absence of vortex pinning on the solid boundaries which is particularly important at low velocities, an order of magnitude smaller than the velocities considered here~\cite{Schwarz1985}.

%%%%%%%%%%%%%%%%%%%%%%%%%%%%%%%%%%%%%%%%%%%%%%%%%%
%%%%%%%%%%%%%%%%%%%%%%%%%%%%%%%%%%%%%%%%%%%%%%%%%%
%%%%%%%%%%%%%%%%%%%%%%%%%%%%%%%%%%%%%%%%%%%%%%%%%%
\begin{figure}
\begin{center}
\includegraphics[width=0.32\textwidth]{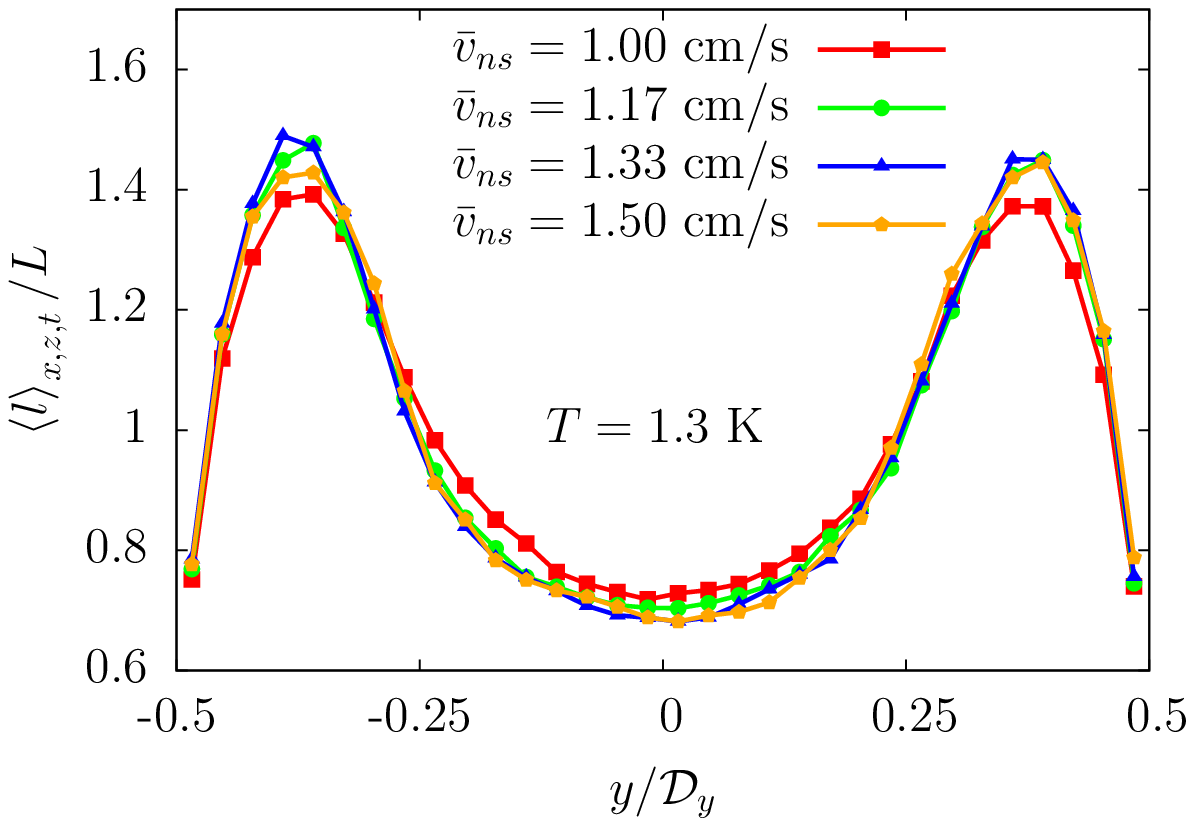}
\hfill
\includegraphics[width=0.32\textwidth]{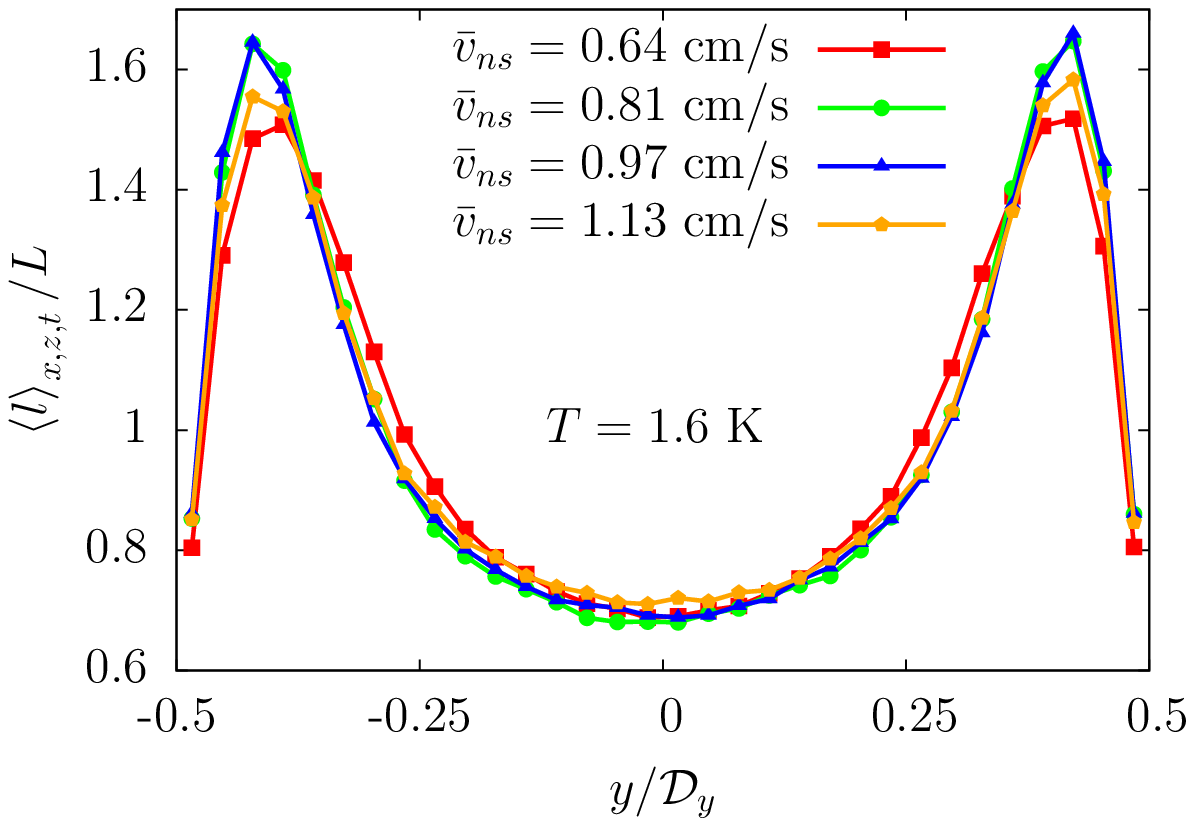}
\hfill
\includegraphics[width=0.32\textwidth]{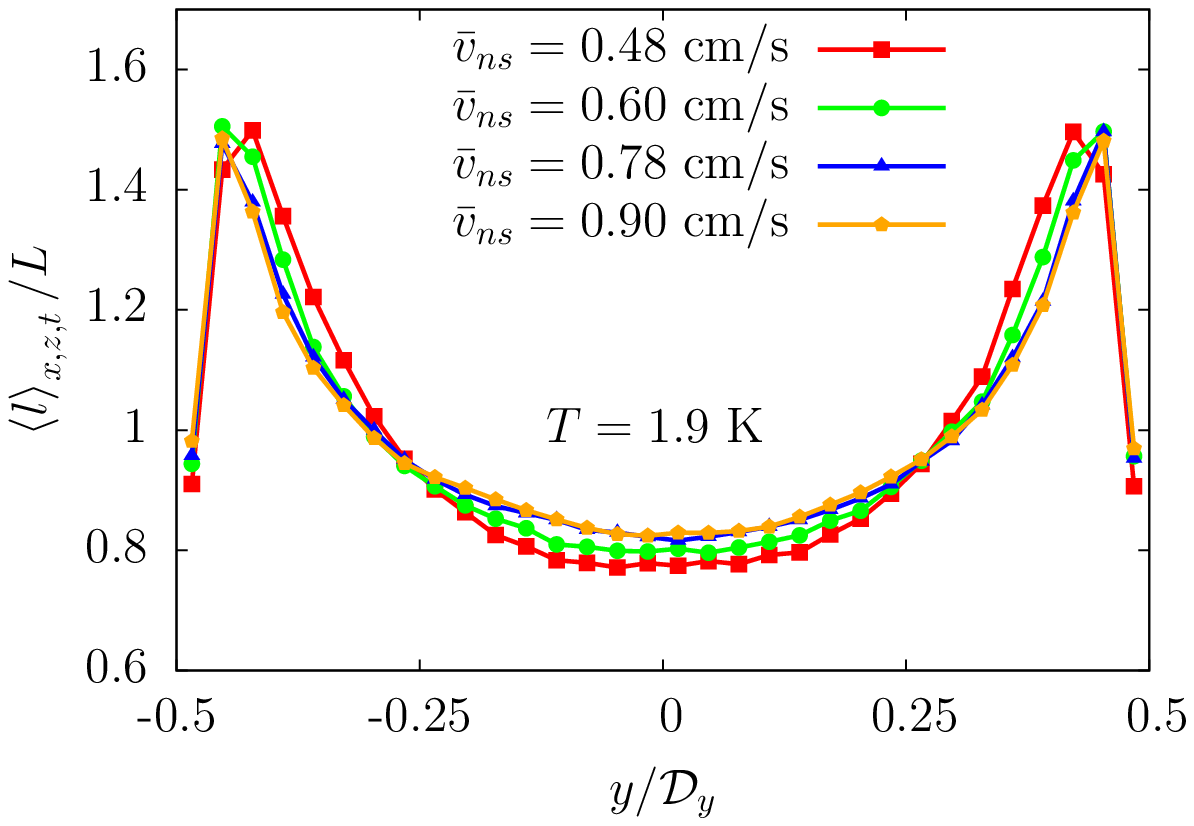}
\caption{Plots of the normalized, $y$-profile of the spatially dependent vortex line density, $\langle l \rangle_{x,z,t}/L$, averaged over time and in the two periodic dimension for each of the three temperatures $T=1.3 ~\rm K$, $T=1.6 ~\rm K$ and $T=1.9 ~\rm K$ respectively.}
\label{fig3}
\end{center}
\end{figure}

%%%%%%%%%%%%%%%%%%%%%%%%%%%%%%%%%%%%%%%%%%%%%%%%%%

\subsection{Coarse-grained statistics across the channel}

%Explain line length
Detailed information about the structure of the tangle is obtained by coarse-graining the vortex tangle onto a Eulerian mesh defined within the computational domain with the spirit of the HVBK approximation in mind~\cite{barenghi2001quantized}. To this end, we define a $M^3=32^3$ uniform cartesian mesh with each volume element, $V_{\rm B}$, a $6.25 \times 10^{-3} ~ {\rm cm} \times 3.125 \times 10^{-3} ~ {\rm cm} \times 3.125\times 10^{-3} ~ {\rm cm}$ cuboid. From the vortex lines, ${\mathcal{L}_{\rm B}}$, which lie within each volume element, $V_{\rm B}$, we can compute estimates of the local vortex line density $l(\bx,t) = (1/{V_{\rm B})\int_{\mathcal{L}_{\rm B}(\bx)}} {\rm d}\xi$. 
%and vortex velocity $\bv=\left\langle {\rm d}\bs/{\rm d}t \right\rangle$  (where angles brackets denote averaging over all segments within the volume) as a function of position $\mathbf{x}$ within each volume element.
 It should be clear from our definitions, that we recover the vortex line density of the whole box by summation over all the sub-volume elements: $L =  \left\langle l\right\rangle_{x,y,z} = \left\langle l\right\rangle_{V} = (1/V) \int_{V} l(\bx,t)\,  {\rm d}x\,{\rm d}y\,{\rm d}z$.

%AWB 5/10/13: cut as repeated below
%To further probe the structure of the tangle vortex length parallel to each of the cartesian directions can be computed,  $L_i(\mathbf{x})=\int_{{\cal L}'} |\bs' \cdot \hat{\mathbf{e}}_i | \, d\xi $, where $i=x,y,z$.

These coarse-grained fields are averaged temporally (over all snapshots of the tangle when the line density has saturated) and then spatially over the two periodic directions $x$ and $z$. We denote the use of this averaging through the notation $\langle\cdot\rangle_{x,z,t}$, leaving quantities purely as a function of position normal to the boundaries, i.e. $\langle l \rangle_{x,z,t}(y)\equiv (1/\mathcal{D}_x\mathcal{D}_z\tau)\int_{t'}^{t'+\tau}\int l(\bx,t) \, {\rm d}x\,{\rm d}z\, {\rm d}t$. 

%%%%Paragraph about vortex line length density
In Fig.~\ref{fig3} we plot the local vortex line length density, normalized by the mean vortex line density of the whole domain, and averaged over both periodic directions, $x$ and $z$, and also over time in steady state conditions.  This gives us the $y$-profile of the distribution of the vortex line length density, namely $\left\langle l\right\rangle_{x,z,t}(y)$.  Fig.~\ref{fig3} displays data from the three different temperatures, $T= 1.3~\rm K$ (left), $T= 1.6~\rm K$ (middle), and $T= 1.9~\rm K$ (right). %Note that in order to best established universality of the line density profile the data has been appropriately normalised by the mean steady state vortex line density $L$.
 Within each subfigure, we plot the profiles for four different velocities $\bar{v}_{ns}$. The higher concentration of vortex line density at the boundaries, visible in Fig.~\ref{fig3}, is immediately obvious. Whilst there is some variation in the actual peak density, for a given temperature and varying $\bar{v}_{ns}$, what is particularly striking is that the {\it location} of the peak is temperature dependent, and appears universal for a given temperature.

\begin{figure}
\begin{center}
\includegraphics[width=0.42\textwidth]{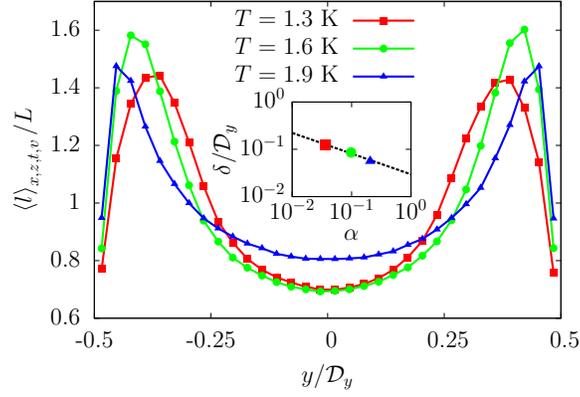}
\caption{Plot of the normalized, spatially dependent vortex line density, $\left\langle l \right\rangle_{x,z,t,v}$, as in Fig.~\ref{fig3}, with an additional average over the velocities $\bar{v}_{ns}$. (Inset) Distance of the peak vortex line density from the wall, $\delta$, scaled by the channel width $\mathcal{D}_y$, plotted as a function of the temperature dependent parameter $\alpha$. A power-law fit corresponding to $\delta/\mathcal{D}_y\propto \alpha^{C}$ is fitted with $C=-0.44$. }%A progressive increase in the width of the boundary layer can be seen with decreasing temperature.}
\label{fig4}
\end{center}
\end{figure}

This observation motivates us to apply further averaging of the data displayed in Fig.~\ref{fig3} over the four velocities.  We depict the resulting vortex line density profile, $\left\langle l \right\rangle_{x,z,t,v}$, in Fig.~\ref{fig4}. We observe a clear concentration of the vortex line density towards the fixed boundaries as temperature increases. Indeed, we plot in the inset of Fig.~\ref{fig4}, the distance of the peak vortex line density from the wall, $\delta$, scaled by the width of the channel ${\cal D}_y $, and compare with the relative value of $\alpha$, the non-dimensional parameter associated to the mutual friction between the normal and superfluid components.  We observe reasonable agreement with the power-law scaling $\delta/{\cal D}_y \propto \alpha^{C}$ with $C\approx -0.5$. The origin of this exponent will be discussed later in the article.
%%%%%%%%%%%%%%%%%%%%%%%%%%%%%%%%%%%%%%%%%%%%%%%%%%
%%%%%%%%%%%%%%%%%%%%%%%%%%%%%%%%%%%%%%%%%%%%%%%%%%
 
%%%%%%%%%%%%%%%%%%%%%%%%%%%%%%%%%%%%%%%%%%%%%%%%%%
\begin{figure}
\begin{center}
\includegraphics[width=0.42\textwidth]{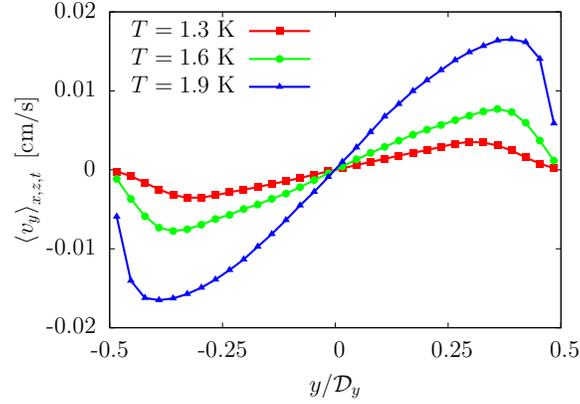}
\caption{Plot of the  $y$-component of the spatially dependent vortex velocity, $\bv$ (cm/s), averaged temporally, spatially (in the two periodic directions $x$ and $z$) and additionally over the velocities (as in Fig.~\ref{fig4}).}
\label{fig5}
\end{center}
\end{figure}
%%%%%%%%%%%%%%%%%%%%%%%%%%%%%%%%%%%%%%%%%%%%%%%%%%
%We also inspect  the isotropy of the vortex tangle, and in particular its dependence on the temperature and the counterflow velocity.  We integrate the normalized tangent $\bs'$ projected on the cartesian basis, in the spirit of Schwarz and others \cite{Schwarz1988, Adachi2010}, who projected the vortex line length in terms of parallel and perpendicular components to the counterflow velocity. Here, due to the effect of the boundaries, we do not expect symmetry between the $y$ and $z$ directions and so project onto all three basis vectors.  Our results are presented in Fig.~\ref{fig7}.  As in previous studies \cite{Adachi2010} we observe that the tangle is anisotropic with the majority of the vortex configuration perpendicular to the counterflow direction. The degree of anisotropy is temperature dependent but independent of the counterflow velocity.   Interestingly we notice a difference in the structure of the tangle, as the temperature is varied. At the highest temperature the vortices are strongly aligned with the $z$ direction, visible in the columnar structures seen in \textcolor{blue}{Fig.~\ref{fig2}} (\textcolor{blue}{far }right). As the temperature is decreased we notice that the length parallel to $y$ and $z$ equilibrate and the tangle is more isotropic with looped structures visible in \textcolor{blue}{Fig.~\ref{fig2}} (\textcolor{blue}{second} left).

To understand this phenomenon, the most obvious quantity to analyze is the $y$-velocity, $v_y$, exhibited by the vortices, which we present (course-grained as above) in Fig.~\ref{fig5}. We observe that $\left\langle v_y\right\rangle_{x,z,t}$ is sinusoidal in profile, meaning that the vortices are being advected towards the solid boundaries.

%%%%%%%%%%%%%%%%%%%%%%%%%%%%%%%%%%%%%%%%%%%%%%%%%%
\begin{figure}
\begin{center}
\includegraphics[width=0.42\textwidth]{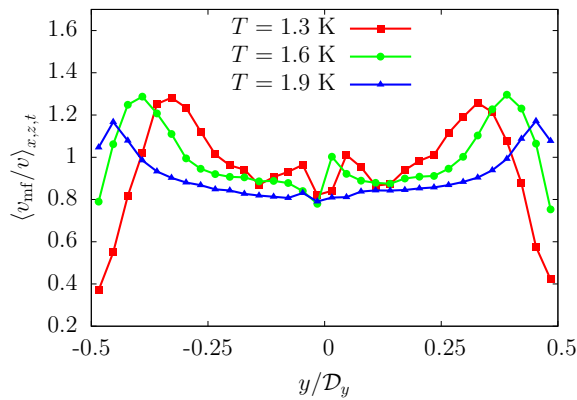}
\caption{ The ratio of the velocity due to mutual friction to the vortex velocity, which is given by $\left\langle v_{\rm mf}/v\right\rangle_{x,z,t}$, where $v_{\rm mf}=|\alpha \bs' \times (\bv_n-\bv_s)
-\alpha' \bs' \times \left[\bs' \times \left(\bv_n-\bv_s\right)\right]|$ and $v=|{\rm d}\bs/{\rm d}t|$.}
\label{fig6}
\end{center}
\end{figure}
%%%%%%%%%%%%%%%%%%%%%%%%%%%%%%%%%%%%%%%%%%%%%%%%%%
Whilst at first blush the figure of the $y$-component of the vortex velocity seems to offer an explanation for the width of the boundary layer, the movement of the peaks of $|v_y|$ to the boundaries with increasing temperature is due to the motion of the boundary layer and not its cause. It is more instructive to consider the ratio of the velocity due to mutual friction and the total velocity, which is given by $v_{\rm mf}/v$, where $v_{\rm mf}=|\alpha \bs' \times (\bv_n-\bv_s)
-\alpha' \bs' \times \left[\bs' \times \left(\bv_n-\bv_s\right)\right]|$. This quantity is displayed in Fig.~\ref{fig6}. It is immediately evident that the peak of this quantity is located close to the maximum of the vortex line density and that within the boundary layer the self induced velocity becomes much more relevant than in the bulk. Hence the question of boundary layer width is simply a balance of two competing effects, the velocity induced by mutual friction pushing the vortices to the boundary, and the self induced velocity which resist this through turbulent diffusion. This balance is evidently temperature dependent, but appears to be independent of the counterflow velocity.

One may also notice in Fig.~\ref{fig4}, the apparent temperature dependence on the width of the vortex line density peaks and the associated flattening of the profile. Initially it may seem inconsistent that at $T=1.9~\rm K$ the resulting turbulence is more homogeneous, than at the lower temperatures. However this is simply the result of a larger external superflow,  due to an increase in the relative density of the normal fluid. As a flat profile is appropriate due to the free-slip boundaries, the greater the contribution $\bv_s^{\rm ext}$ makes to $\bar{v}_{ns}$, the more homogeneous the resulting vortex tangle will be. 

\begin{figure}
\begin{center}
\includegraphics[width=0.42\textwidth]{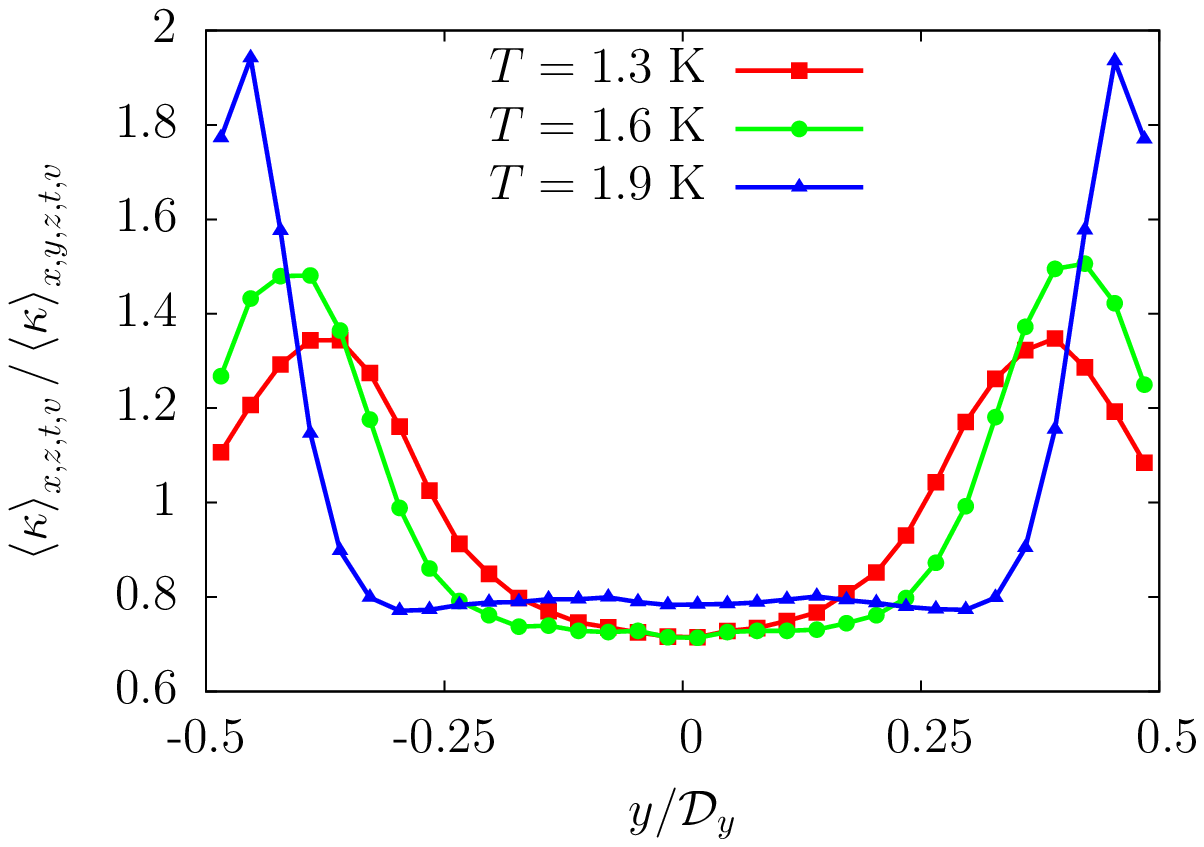}
\includegraphics[width=0.42\textwidth]{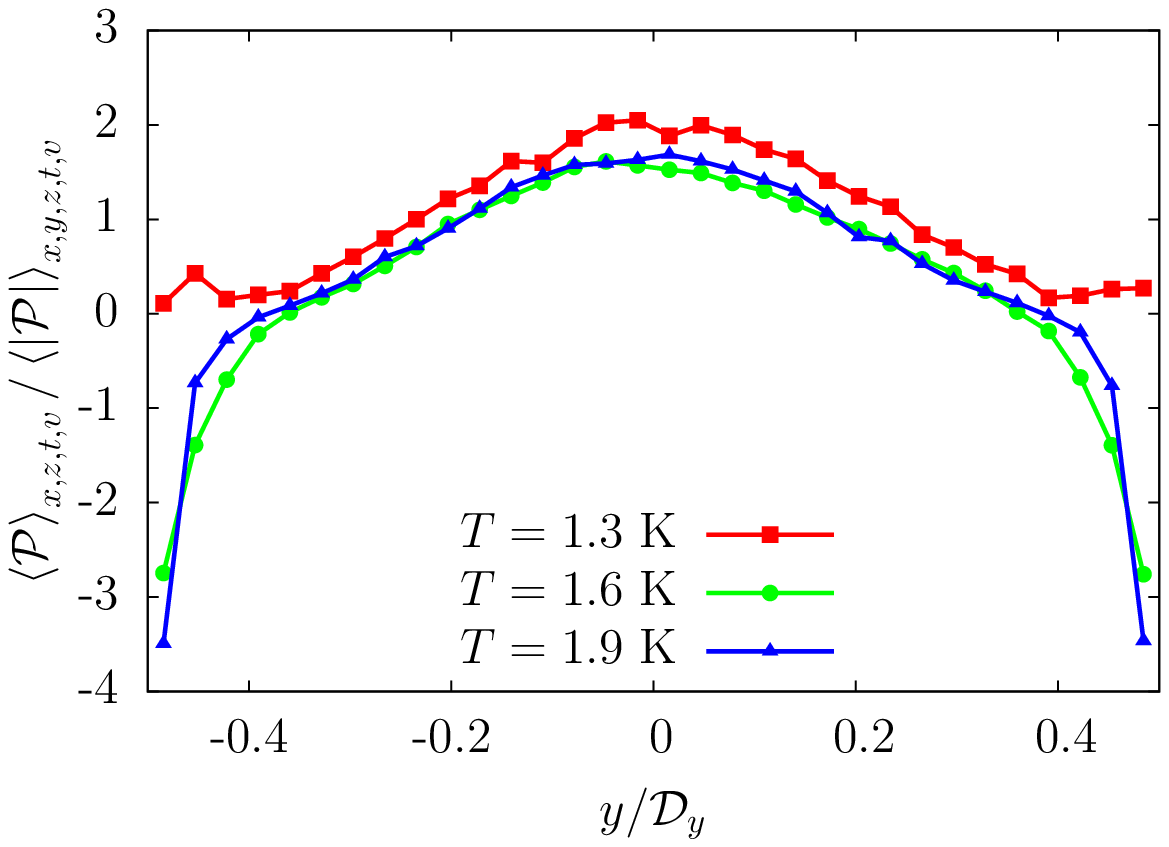}
\caption{(left) The coarse-grained curvature of the vortex filaments $\kappa=|\bs''|$ along the channel; (right) the coarse-grained local stretching rate $\cal P=-\kappa \bv \cdot \mathbf{n}$ along the channel.}
\label{newfig1}
\end{center}
\end{figure}

Given the peaks in the vortex line density it is also instructive to analyse regions of quantized vorticity production. The rate of change of the vortex line length in our system can be expressed as
\begin{equation}
\frac{{\rm d}\Lambda}{{\rm d}t}=\frac{{\rm d}}{{\rm d}t}\oint_{\mathcal{L}} |\bs'| \,{\rm d}\xi.
\end{equation}
After a little manipulation this can be shown to be equivalent to
\begin{equation}
\frac{{\rm d}\Lambda}{{\rm d}t}=-\oint_{\mathcal{L}} \kappa \bv \cdot \mathbf{n} \, {\rm d}\xi,
\end{equation}
were $\kappa(\xi,t)=|\bs''|$ is the curvature of the vortex segment, $\bv(\xi,t)$ the segments velocity and $\mathbf{n}(\xi,t)$ the local normal vector to the vortex line. Hence, we define ${\cal P}(\xi,t)=-\kappa \bv \cdot \mathbf{n} $ as the local vorticity production rate; as above this is course-grained and appropriately averaged to leave it as a function of the wall normal direction only. Clearly from this expression, both the velocity of the vortices and the curvature are crucial quantities for determining vortex stretching or contraction. It is therefore interesting to plot the course-grained and averaged curvature profile $\kappa(y)$ for the three temperatures across the channel, see Fig.~\ref{newfig1} (left). It is clear that the curvature is much larger in the near the solid boundaries, most likely due to the larger vortex density and reconnection rate (see Fig. \ref{newfig3}). However when we turn our attention to the stretching rate, Fig.~\ref{newfig1} (right), we see that this is larger in the centre of the channel, and for the two higher temperatures becomes negative close to the boundaries where the curvature is large.
%%%%%%%%%%%%%%%%%%%%%%%%%%%%%%%%%%%%%%%%%%%%%%%%%%
%%%%%%%%%%%%%%%%%%%%%%%%%%%%%%%%%%%%%%%%%%%%%%%%%%

This result can be better understood if we consider the isotropy of the tangle across the channel. We compute the local vortex length parallel to the counterflow direction as $\Lambda_{x}(\bx,t) = \int_{\mathcal{L}_{\rm B}(\bx)} \bs' \cdot \mathbf{e}_x\, {\rm d}\xi$. It is well known~\cite{Schwarz1988,Adachi2010} that counterflow turbulence is anisotropic, with vortices tending to be oriented perpendicular to the counterflow direction, this effect is more pronounced with increasing mutual friction.  Fig.~\ref{newfig2} shows the course-grained and averaged parallel length $\left\langle \Lambda_x \right \rangle_{x,z,t,v}$, normalized by the total vortex line length $\left\langle \Lambda\right \rangle_{x,z,t,v}$. We see that the tangle is indeed anisotropic ($\Lambda_x/\Lambda=1/\sqrt{3}$ is a necessary but not sufficient condition for the tangle to be isotropic) with increasing anisotropy as the temperature is increased. However a marked increase in the isotropy of the tangle is seen as one moves towards the boundaries. From this result we can surmise that in counterflow turbulence the quantized vortices are very organised in the centre of the channel and expand due to the mutual friction force. However within the boundary layer region close to the solid boundaries the tangle is much more random, and so the effect of mutual friction is to contract the vortices, as one would observe for a single vortex ring propagating within a stationary normal fluid. 

\subsection{Analysis of the vortex line density profile}

\begin{figure}
\begin{center}
\includegraphics[width=0.42\textwidth]{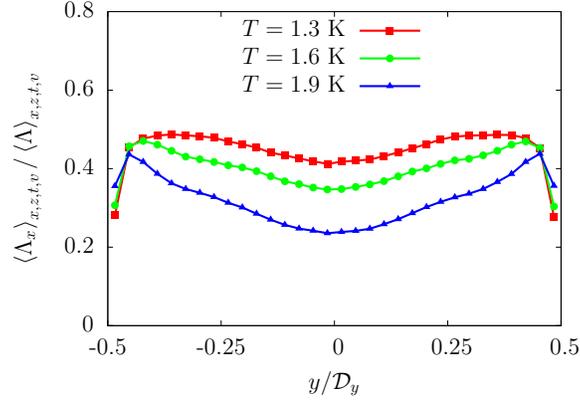}
\caption{Normalized vortex line length parallel to the counterflow direction, scaled by the total vortex length. Note the strong anisotropy in the centre of the channel, particularly for $T=1.6~\mathrm{K}$ and $T=1.9~\mathrm{K}$, and the increase in the isotropy  as we move towards the solid boundaries.}
\label{newfig2}
\end{center}
\end{figure}

This insight into the structure of the tangle prompts us to revisit the power law relationship between the distance of the peak vortex density from the wall and the mutual friction parameter $\alpha$.
Our picture is the following, the mutual friction generates a very anisotropic tangle in the centre of the channel where the normal fluid velocity is largest. Kelvin wave perturbations along the vortices are amplified by the Donnelly-Glaberson instability and $\Lambda_x<\Lambda_{y},\Lambda_{z}$ hence $v^{si}_y<v^{si}_x$. This is evident from the positive quantized-vorticity rate seen in the centre of the channel in Fig.~\ref{newfig1}, and the anisotropy of the tangle visible in Fig.~\ref{newfig2}. In essence the quantized vortices are slave to the mutual friction and are pushed out towards the solid boundaries. Subsequently they arrive at the edge and reconnect, creating a random tangle at the boundary. Subsequent vortices arrive at the boundary which reconnect with vortices at the walls creating a turbulent boundary layer where the vortex density clearly peaks in our simulations.
As we observe, the structure of the tangle is very different in centre compared to that at the boundaries, with these boundary layer regions being areas of approximately homogeneous isotropic turbulence.  This motivates us to try and predict the location of the vortex line density peaks by considering the balance of two competing mechanisms.   Close to the boundaries, we have an almost random tangle that will diffuse over a distance $d$  with the rate given by
\begin{equation}
\tau_{\rm diff} \sim \Gamma/d^2.
\end{equation}
It is well known that the energy from the normal fluid flow is transferred to quantized vortices via the Donnelly-Glaberson instability, where Kelvin waves of a given wave number $k$ are destabilized with a growth rate \cite{Ostermeier1975,Tsubota:2004}
\begin{equation}
\tau_{\rm DG} \sim \alpha(kv_{ns}-\nu' k^2).
\end{equation}
Here $\nu'=(\Gamma/4\pi)\ln(1/ka_0) \approx \Gamma$;
this is also the timescale at which vortices are pushed to the boundaries. If we seek the balance of these two competing effects then this would provide a natural length scale for the width of the boundary layer:
\begin{equation}
\label{BL_width}
\delta\sim\sqrt{\frac{\Gamma}{\alpha(kv_{ns}-\nu' k^2)}}.
\end{equation} 
Notice, that the dependence to the mutual friction coefficient gives $\delta \sim \alpha^{-1/2}$, consistent with the observed scaling from our numerical simulations, $\delta \sim \alpha^{-0.44}$.

Earlier we argued that there appeared to be little dependence on the counterflow velocity at a given temperature, whilst in Eq.~(\ref{BL_width}) $v_{ns}$ appears explicitly.  Whilst the mean counterflow velocity varies by approximately 50\% for the set of simulations performed at a given temperature, the counterflow velocity close to the boundary layer will vary by much smaller amounts due to the parabolic profile of the normal fluid velocity. Finally if we take the data from one of the simulations at $T=1.6~\mathrm{K}$ ($\alpha \simeq 0.1$) with $\bar{v}_{ns} \simeq 1$cm/s we find the inter-vortex spacing $\ell \simeq 0.125~\mathrm{cm}$. If we take a wavenumber for the Donnelly-Glaberson growth rate based on this length scale, i.e. $k_\ell=2\pi/\ell \simeq 500 \, {\rm cm}^{-1}$ and substitute these numbers into Eq.~(\ref{BL_width}) we estimate that $\delta \simeq 0.006~\mathrm{cm}$. This is in reasonable agreement with the peak of the density which occurs around $0.01~\mathrm{cm}$ from the boundary at $T=1.6~\mathrm{K}$, particularly given that our estimate should provide a lower bound on the width as we are using the mean counterflow velocity in our calculation.

%%%%%%%%%%%%%%%%%%%%%%%%%%%%%%%%%%%%%%%%%%%%%%%%%%
%%%%%%%%%%%%%%%%%%%%%%%%%%%%%%%%%%%%%%%%%%%%%%%%%%

The results presented so far are valid in systems where the normal fluid is laminar (TI regime). Above a critical heat flux, a further transition in the system is observed, with a dramatic increase in the observed values of $\gamma$~\cite{Tough1982}; this TII regime is typically associated with a transition to turbulence in the normal fluid.

\begin{figure}
\begin{center}
\includegraphics[width=0.42\textwidth]{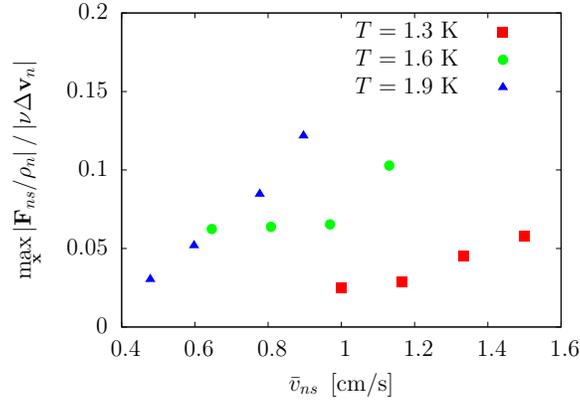}
\caption{Ratio of the maximum attained values of the mutual friction term ${\bf F}_{ns}$ given by Eq.~(\ref{eq:mutualfriction}) over the magnitude of the viscous term in Eq.~(\ref{eq:NavierStokes}) given the Poiseuille normal fluid profile which results in $\left| \nu \Delta \bv_n \right|=2\nu U_c/h^2=3\nu\bar{v}_{ns}\rho_s/h^2\left(\rho_s+\rho_n\right)$.}
\label{fig7}
\end{center}
\end{figure}

To verify our assumptions that our simulations do indeed correspond to the TI regime, we estimate the maximum contribution of the mutual friction term ${\bf F}_{ns}$ in the Navier-Stokes equation given by Eq.~(\ref{eq:mutualfriction}). We numerically perform the line integral, determining the values of the mutual friction term at points where the quantized vortices exist.  We record the maximum attained value of $\left|{\bf F}_{ns}\right|$ along the vortex tangle and compute the ratio over the contribution arising through the viscous term, which in the case for the Poiseuille profile is spatially independent (constant) over the whole domain. In Fig.~\ref{fig7}, we present the ratio of the maximum of $\left|{\bf F}_{ns}\right|$ divided by the absolute value of the viscous term. We observe that the mutual friction is at least a magnitude smaller that the viscous term in all our simulations.  Thus, viscous effects dominate in the Navier-Stokes equations, and so the normal fluid flow will not substantially feel any mutual friction effects resulting from the drag of the quantized vortex tangle of the superfluid component.  Ultimately, with the negligible mutual friction contribution and the low Reynolds numbers, we conclude that the normal fluid flow will not be destabilized from its laminar state (TI) by the presence of the quantum tangle and our assumption of neglecting the back-reaction of the superfluid onto the normal fluid is justified.

%%%%%%%%%%%%%%%%%%%%%%%%%%%%%%%%%%%%%%%%%%%%%%%%%%
%%%%%%%%%%%%%%%%%%%%%%%%%%%%%%%%%%%%%%%%%%%%%%%%%%

\subsection{Reconnection statistics}

Before we conclude our analysis of the TI regime, we analyse the reconnection statistics. Reconnections are important events, changing the topology of the vortex tangle, and dissipating kinetic energy which allows the system to be driven into a non-equilibrium steady state. In a series of papers~\cite{Poole2003,BarenghiRecon04} Barenghi, Samuels and coauthors considered how the reconnection rate $\zeta$ (defined as the number of vortex reconnections per unit time per unit volume) depends on the vortex line density $L$. They argued, and provided numerical evidence, for two regimes, both with a power-law relationship, $\zeta \sim L^\beta$. For a homogeneous, isotropic tangle of vortices Poole \etal~\cite{Poole2003} provided a simple geometric argument which leads to a scaling with $\beta=5/2$. In contrast, if the mutual friction term in Eq.~(\ref{eq:mutualfriction}) dominates over the self induced term $\bv_s^{\rm si}$, the same geometric arguments lead to a scaling with $\beta=2$. Note both arguments are based on the assumptions of homogeneity and isotropy, which is clearly not valid for the tangles generated in our numerical simulations. However in numerical simulations of counterflow, which were homogeneous but anisotropic, Barenghi and Samuels \cite{BarenghiRecon04} showed reasonable agreement with the $\beta=2$, although with some variation in the scaling of their numerical simulations.

\begin{figure}
\begin{center}
\includegraphics[width=0.42\textwidth]{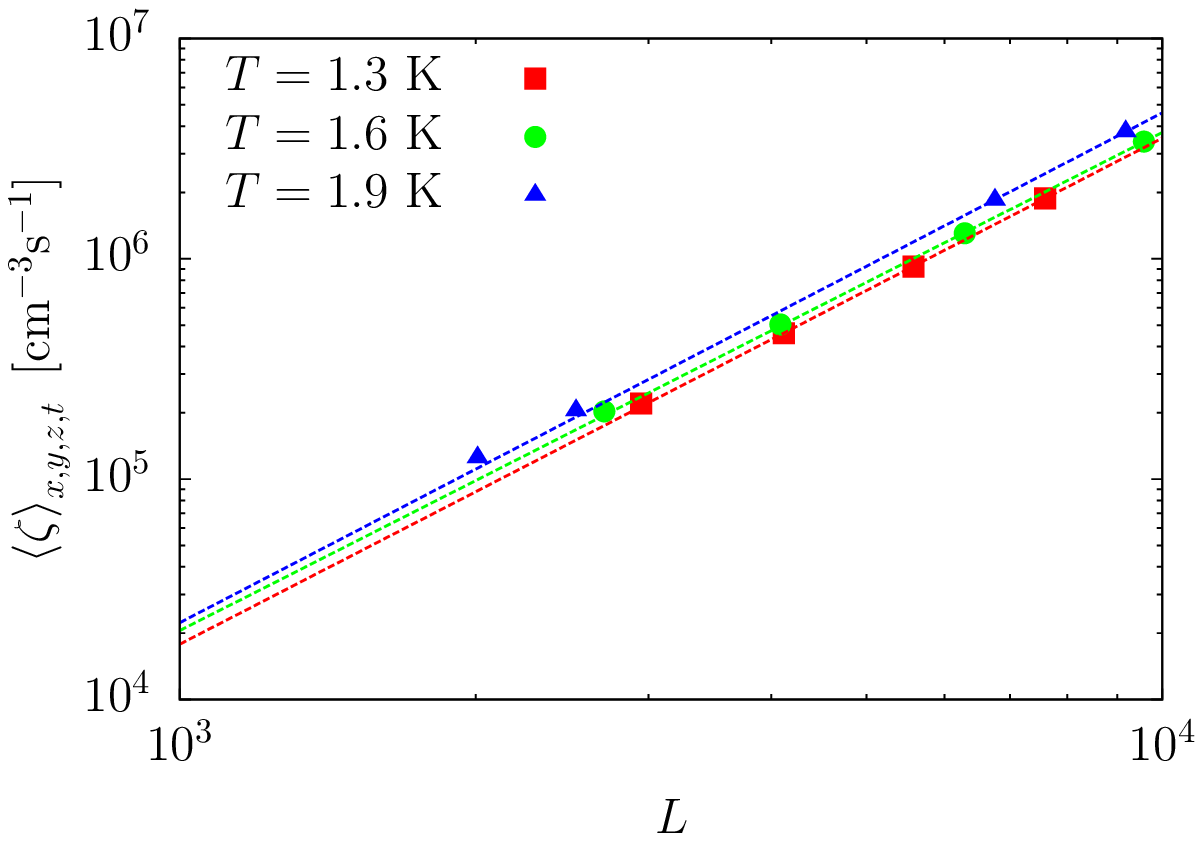}
\includegraphics[width=0.42\textwidth]{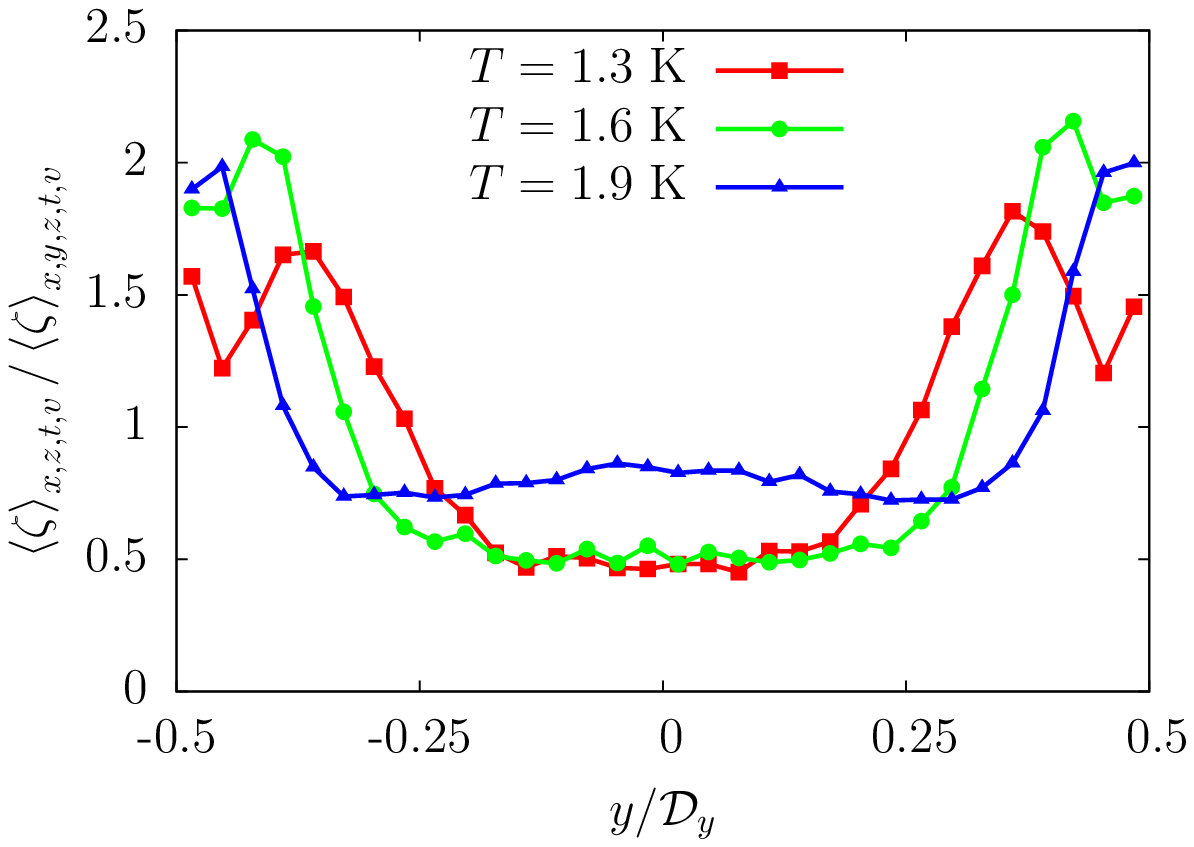}
\caption{(left) The total reconnection rate $\zeta$  the total vortex line density $L$. The dashed straight lines indicated the best power-law fit with scalings $\zeta \sim L^{2.29}$ ($T=1.3~\mathrm{K}$), $\zeta \sim L^{2.26}$ ($T=1.6~\mathrm{K}$) and $\zeta \sim L^{2.31}$ ($T=1.9~\mathrm{K}$). (right) The coarse-grained profile along the channel of the reconnection rate.}
\label{newfig3}
\end{center}
\end{figure}

\begin{figure}
\begin{center}
\includegraphics[width=0.32\textwidth]{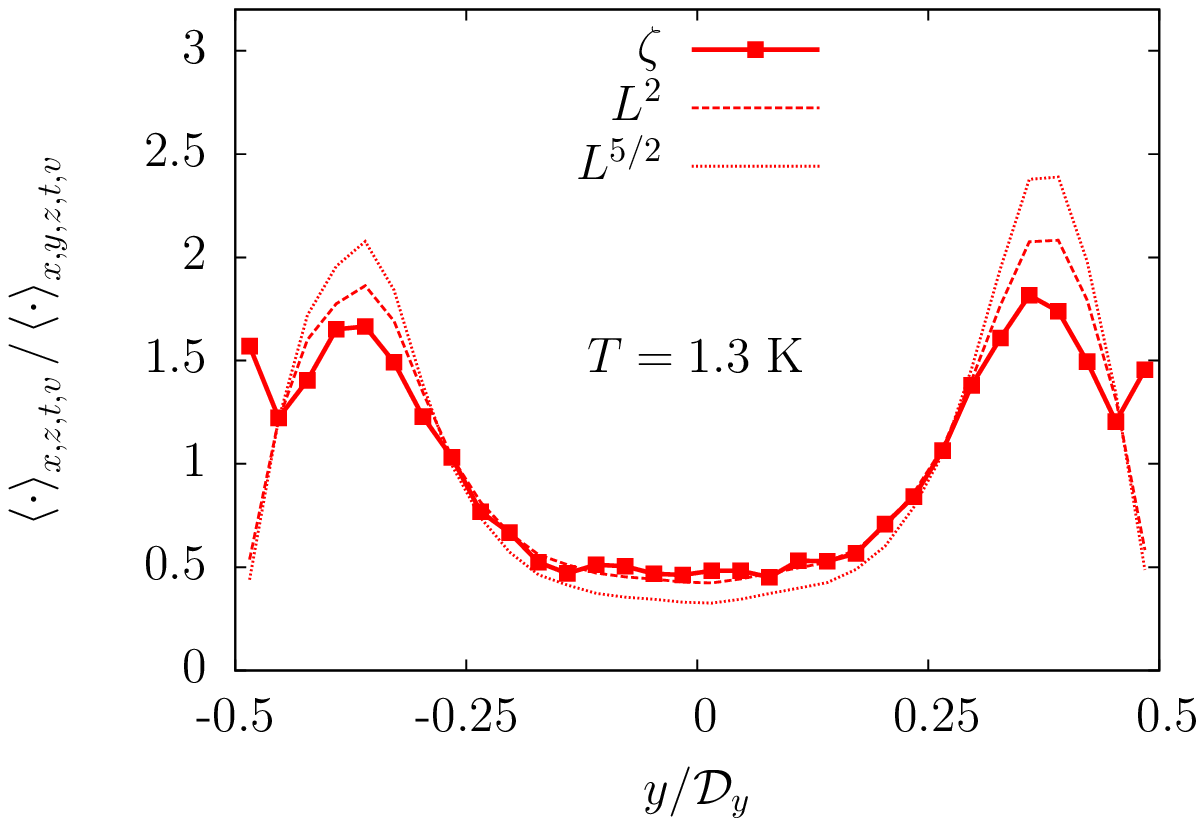}
\includegraphics[width=0.32\textwidth]{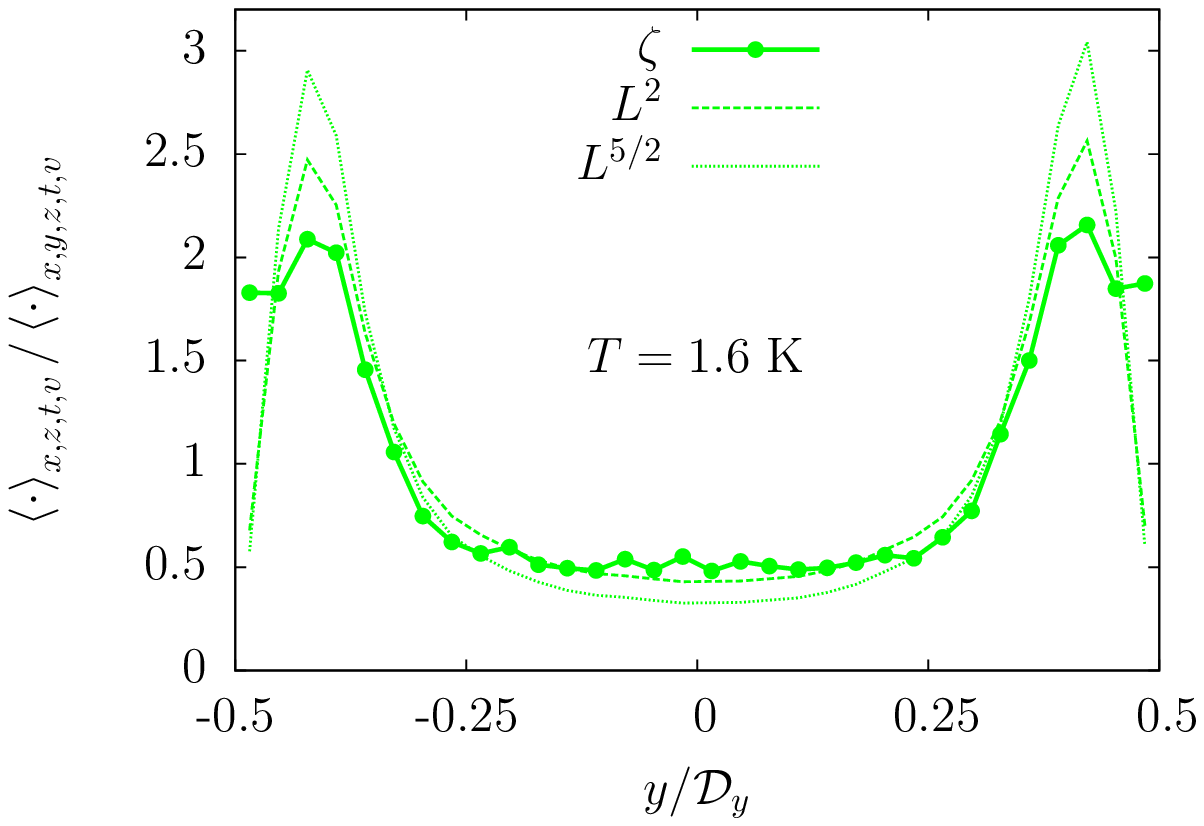}
\includegraphics[width=0.32\textwidth]{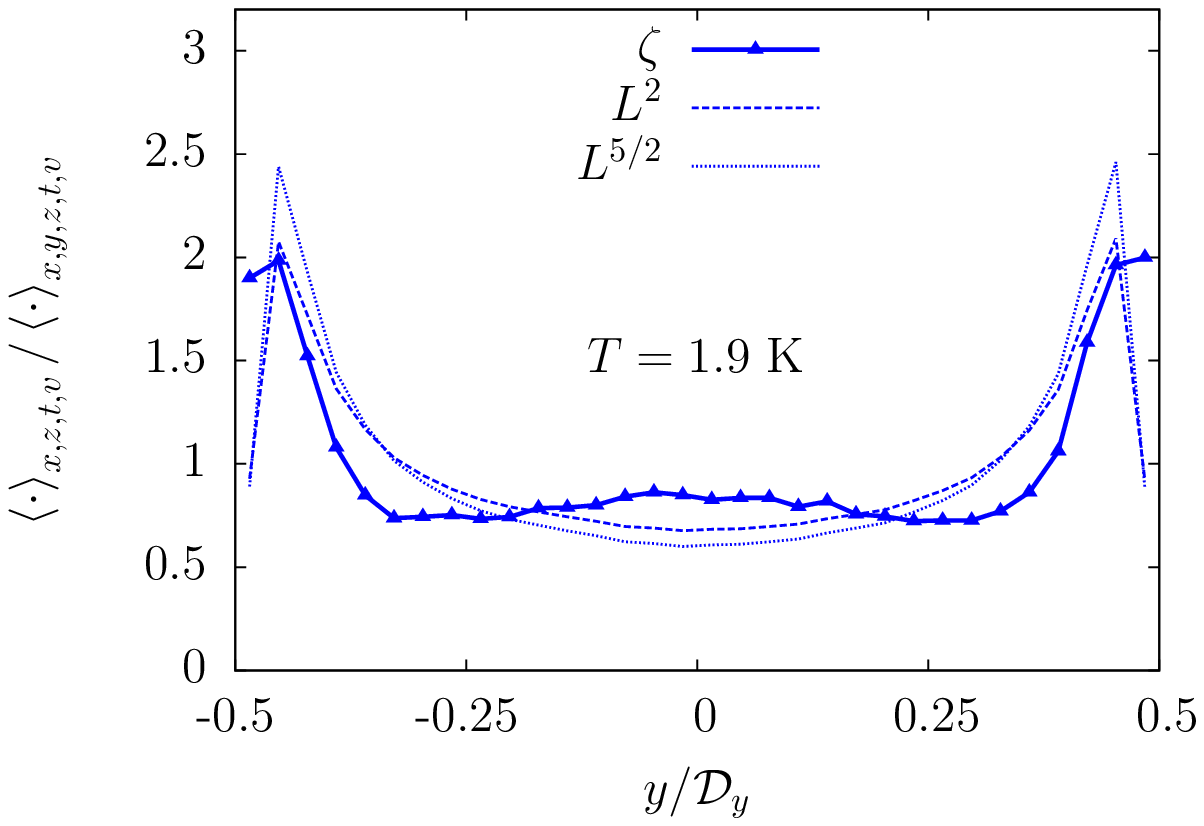}
\caption{Comparison of the theoretical predictions of \cite{Poole2003,BarenghiRecon04} $\zeta \sim L^{\gamma}$, $\gamma=5/2$ (zero temperature, random isotropic tangle), $\gamma=2$ (high temperature, TI regime) with the data from our numerical simulations.}
\label{newfig4}
\end{center}
\end{figure}

Fig. \ref{newfig3} (left) shows the reconnection rate verses the vortex line density for our data. We find a power law scaling, $\zeta \sim L^\beta$, with $\beta \approx 2.3$, almost independent of the mutual friction parameters. Note in determining the reconnection rate we have only considered reconnections between vortices, and not reconnections between vortices and the solid boundaries.  However, reconnections between vortices and the solid boundaries are likely to become important in the limit $T\to 0$~\cite{Eltsov2010a,Eltsov2010b}. Clearly this exponent lies between the two values discussed above. In the right panel of Fig. \ref{newfig3} we show the spatial dependence on the reconnection rate across the channel. As we expect, the rate is much larger close to the boundaries where the vortex line density is concentrated. Of course a natural question one may ask is to what extent can the spatial dependence of the reconnection rate be explained by the theoretical predictions of \cite{Poole2003,BarenghiRecon04}, using the course-grained vortex line density profiles? Fig. \ref{newfig4} displays the reconnections rates from Fig. \ref{newfig3} alongside the `predicted' reconnection rates, based on the course-grained vortex line densities. A number of interesting features are apparent, in the centre of the channel the $L^2$ scaling gives a reasonable description of the reconnection rate, particularly for the two lower temperatures  $T=1.3~\mathrm{K}$ and $T=1.6~\mathrm{K}$. At all temperatures, within the boundary layer region, we see a relative suppression of the reconnection rate, at least from what we would expect based on the vortex line density in Fig.~\ref{fig4}. This is a little surprising given that as this region in vortex tangle shows a greater degree of isotropy we would perhaps expect the $5/2$ exponent to provide a good fit; this result warrants further investigation in the future. Finally we note that very close to the boundaries, where the vortex line density is dramatically reduced, the reconnection rate remains anomalously large. We have verified using smaller bin widths (and hence a larger number of mesh points) that this is not an artefact of our course-graining procedure. This relatively large number of vortex reconnections so close to the boundary, could be important in understanding the TI to TII transition. Indeed the classical mechanism for the destabilisation of boundary layer flow is due to the Tollmien-Schlichting instability~\cite{charru2011hydrodynamic}, which arises in the viscous boundary layer close to the wall.

%%%%%%%%%%%%%%%%%%%%%%%%%%%%%%%%%%%%%%%%%%%%%%%%%%
%%%%%%%%%%%%%%%%%%%%%%%%%%%%%%%%%%%%%%%%%%%%%%%%%%

%%%%%%%%%%%%%%%%%%%%%%%%%%%%%%%%%%%%%%%%%%%%%%%%%%
%%%%%%%%%%%%%%%%%%%%%%%%%%%%%%%%%%%%%%%%%%%%%%%%%%

\subsection{The TI to TII transition}

Having examined the effect of a laminar normal fluid, an interesting question is to what extent these results carry into the TII regime, thought to occur due to the onset of turbulence in the normal fluid component. Hence we briefly revisit the work of~\cite{BaggaleyLaizet} and extend their results to include the three temperatures considered here.
Our goal is solely to investigate the extent to which an imposed velocity field with some of the properties of turbulent channel flow changes the profile of the vortex line density, and if applicable the power-law scaling presented in Eq.~(\ref{BL_width}). To this end we replace the imposed parabolic normal fluid profile with the snapshot of a Direct Numerical Simulation (DNS) of turbulent channel flow, generated from numerically solving the Navier-Stokes equation in a similar periodic channel geometry as used in this study. Note the statistical properties of this velocity field have been benchmarked against the literature~\cite{BaggaleyLaizet}.

\begin{figure}
\begin{center}
\includegraphics[height=4.1cm]{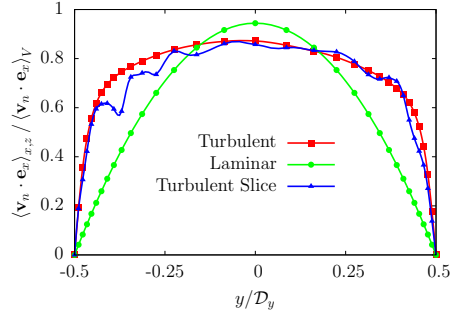}
\caption{Normalized $y$-profile of the averaged normal fluid velocity, taken from the DNS and compared with that of a laminar Poiseuille flow given by Eq.~(\ref{eq:profile}) and a single slice of the turbulent snapshot. \label{fig9} }
\end{center}
\end{figure}

\begin{figure}
\begin{center}
\includegraphics[width=0.42\textwidth]{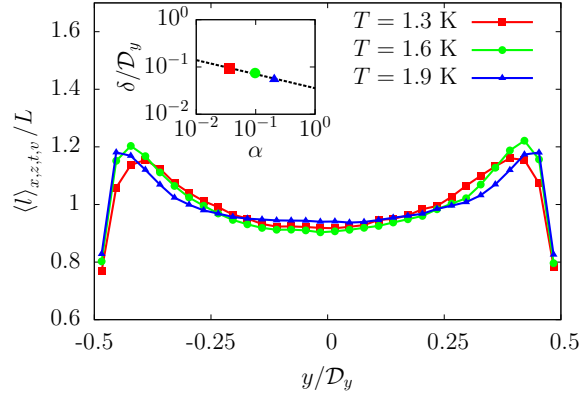}
\caption{Plot of the normalized, $y$-profile of the spatially dependent vortex line density $\left\langle l\right\rangle_{x,z,t}/L$, averaged over the two periodic directions for the vortex tangle generated by the turbulent normal fluid profile.  (inset) Normalized distance of the peak vortex line density from the wall, $\delta/\mathcal{D}_y$, plotted as a function of the temperature dependent parameter $\alpha$. A power-law fit corresponding to $\delta/\mathcal{D}_y\propto \alpha^{C}$ is fitted with $C=-0.30$. }
\label{fig10}
\end{center}
\end{figure}

 Due to the dimensions of the frozen DNS snapshot that we use, we now consider a domain size $\mathcal{D}_x \times \mathcal{D}_y \times \mathcal{D}_z = 0.63~{\rm cm}\times 0.1~{\rm cm}\times 0.21~{\rm cm}$.  We scale the DNS velocity such that $\bar{v}_{ns}=1.17~\rm cm/s$ ($T=1.3~\rm K$), $\bar{v}_{ns}=0.664~\rm cm/s$ ($T=1.6~\rm K$), and $\bar{v}_{ns}=0.577~\rm cm/s$ ($T=1.9~\rm K$), which result in comparable vortex line densities.  We note, that by doing this, the velocity field is no longer a solution to the Navier-Stokes equations, but still gives a flattened profile qualitatively similar to classical turbulent channel flow, see Fig.~\ref{fig9}.  We remark that the transition from a laminar to turbulent normal fluid may arise through instabilities from the nonlinear term of the Navier-Stokes equations (high Reynolds number) or through perturbations resulting from the presence of the quantum turbulence tangle (mutual friction with the superfluid component). The simulations are time-stepped as before, using the new aspect ratio, until the vortex line density saturates and then the coarse-grained vortex line length density is computed in Fig.~\ref{fig10}. We immediately notice that the structure of the tangle is much more homogeneous. However, the peaks of the vortex line density are still  present, although not quite as pronounced as in the laminar simulations.  From the inset of Fig.~\ref{fig10}, we observe that the position of the peaks across the channel are still temperature dependent, with the location of the peaks moving towards the boundaries with increasing temperature with a comparable power-law scaling to that observed in the laminar case, with the predicted $\alpha^{-1/2}$ scaling. We do however note that the discrepancy between our theoretical prediction and the computational data is larger than for the laminar (TI) simulations, and this certainly warrants revisiting in a more realistic study with full coupling between the two fluid components.

The transition to turbulence in the normal fluid has received a large amount of experimental interest recently, most notably from the work of Guo and collaborators~\cite{Guo2010,Guo2013}. They have made use of He$_2$ excimer molecules as tracers to effectively paint a fluorescent strip in the normal fluid. The deformation of this strip can then be observed and the state of the normal fluid inferred. Clearly this is one of the most promising techniques to carefully probe the TI-TII transition, however our results here perhaps indicate that older and more established methods could also be used to indirectly ascertain the state of the normal fluid. The most common technique to estimate the vortex line density in superfluid turbulence is through the application of second sound probes~\cite{Babuin12}. In brief, second sound (temperature) waves can be generated on one side of the channel and their amplitude, after traversing the channel, can be measured on the opposite side. Second sound is attenuated due to the scattering of thermal excitations (which constitute the normal fluid) by quantized vortices; from measuring the reduction in amplitude of the second sound signal when the heater is switched on, the vortex line density can be estimated.  With higher harmonics of the second sound, it is possible, in principle, to crudely measure the structure of the tangle across the channel~\cite{Skrbekpriv}. Given the results presented here, it is reasonable to consider that the transition to the turbulent normal fluid state could be detected, without directly imaging the flow, but through monitoring of the homogeneity of the vortex tangle. The present authors are not aware that such a suggestion has previously been made in the literature.

\section{Conclusions}

We have investigated boundary effects on superfluid counterflow turbulence at finite temperature in a system with fixed boundary conditions.  We considered, and ultimately justified, the use of a Poiseuille flow profile for the normal fluid velocity and used the Schwarz equations to model the evolution of the superfluid component.  We showed that the well-known relationship between vortex line density and the counterflow velocity is in agreement with previous studies whilst also investigating the effects of no-slip boundaries.  We found that the vortex line density concentrates close to the solid boundaries where the majority of the vortex reconnections occur.  In this region, we find that the mean curvature is relatively large with the tangle almost isotropic. Conversely, in the centre of the channel reconnections are suppressed due to the anisotropy of the tangle with the majority of the vortices being oriented perpendicular to the normal fluid flow. Consequently, the centre of the channel is also the region of quantized-vorticity production. Moreover, we find that the position of the vortex line density peaks gradually migrate towards the boundaries with increasing temperature. We show that the numerical data of this position can be explained by a theoretical prediction based upon the balance of turbulent diffusion at the walls and quantized-vorticity production in the centre. The same effect is also observed when we consider a frozen snapshot of DNS turbulence as a model for a turbulent normal fluid, however with a noticeable increase in the homogeneity of the tangle in this case. This observed increase of homogeneity in the tangle could be used as an indirect indication of the transition to turbulence in the normal fluid component.

\begin{acknowledgements}
We would like to acknowledge Jeremie Bec and Risto H\"anninen for helpful discussions. This work was supported by the Carnegie Trust.
\end{acknowledgements}

\bibliographystyle{spphys}       % APS-like style for physics

\end{document}